# Rethinking Activity Awareness: The Design, Evaluation & Implication of Integrating Activity Awareness into Mobile Messaging

Ling Chen* and Miaomiao Dong

*College of Computer Science and Technology, Zhejiang University, Hangzhou, China*

*corresponding author: lingchen@cs.zju.edu.cn, College of Computer Science and Technology, Zhejiang University, 38 Zheda Road, Hangzhou 310027, China.

The biographies of all authors:

**Ling Chen** is a Professor in the College of Computer Science and Technology, Zhejiang University, China, where he received his Ph.D. in computer science in 2004. His research interests include data mining, ubiquitous computing, and human-computer interaction.

**Miaomiao Dong** is a Ph.D. student in the College of Computer Science and Technology, Zhejiang University, China. Her research interests mainly focus on human-computer interaction.

# Rethinking Activity Awareness: The Design, Evaluation & Implication of Integrating Activity Awareness into Mobile Messaging

Nowadays, different types of context information are integrated into mobile messaging to increase expressiveness and awareness, including mobile device setting, location, activity, and heart rate. Due to low recognition accuracy, sometimes users cannot accurately infer others' status through activity awareness. Recently, activity recognition technology has advanced. However, the user behaviors of activity awareness with improved technology have not been studied. In this study, we design ActAware, a mobile instant messaging application that integrates activity awareness based on improved activity recognition technology, i.e., improved recognition accuracy and the addition of activity transition notification. We conduct a field study to explore user behaviors and found that activity awareness allows users to speculate on the reasons for chat interruption, plan communication, speculate on whether the chat partner is departing/arriving, and deepen the understanding of living patterns. Compared with disclosing other types of context information, users have fewer privacy concerns about disclosing activity information in ActAware. Based on these findings, we provide design recommendations for mobile messaging to better support activity awareness.

Keywords: Activity awareness; context information; mobile messaging; design; evaluation

## 1. Introduction

With the popularity of mobile devices, mobile messaging is becoming one of the most important communication channels (Bodic, 2005; Church & Oliveira, 2013; Ardèvol-Abreu et al., 2022), which includes short message service and mobile instant messaging. Nowadays, in order to increase expressiveness and awareness, different types of context information are integrated into mobile messaging (Oulasvirta et al., 2005; Bentley & Metcalf, 2007; Hassib et al., 2017; Griggio et al., 2019), including mobile device setting, location, activity, and heart rate. Researchers have designed mobile messaging applications that integrate context awareness to explore user behaviors and design dimensions (Consolvo et al., 2005; Oulasvirta, 2008; Liu et al., 2021). They have found that context awareness could help users understand each others' status, enhance communication, and shorten the distance between each other, but users had privacy concerns about disclosing their location and mobile device settings.

Due to technology limitations, only a few studies have explored activity awareness in mobile messaging. Early activity recognition technology could only obtain users' motion information, i.e., *moving* or *not moving*. Through a field study, researchers have found that users could use motion information to plan communication, coordinate in-person get-togethers, and stay connected in each others' lives (Bentley & Metcalf, 2007). In addition, users did not worry about privacy problems and intended to obtain more activity information about others. Recently, activity recognition technology has aided in obtaining more detailed activity information, e.g., *walking*, *running*, and *in vehicle*. Researchers have used this technology to develop a mobile messaging application that integrates activity awareness and have studied user behaviors (Buschek et al., 2018). This application only allows users to view each others' activities when sending messages, and the activity information access control is limited. With the addition of low recognition accuracy, sometimes users could not accurately infer the status of others, and they assumed that the activity is the least interesting context in mobile messaging. At present, activity recognition technology has further advanced (Chen et al., 2020; Chen et al., 2021). In 2020, Google Play Services upgraded Activity Recognition API to Activity Recognition Client[1], which improves accuracy, consumes less power, and enhances engineering productivity. In addition to improving the recognition accuracy, Activity Recognition Client also developed Activity Recognition Transition API, which can be used to detect users' activity transitions. However, the application design and user behaviors of activity transition have not been studied. With the improvement of recognition accuracy, mobile messaging applications that integrate activity awareness can obtain more accurate activity information of users. Whether this information raises users' privacy concerns and what are users' motivations for using this information have not been studied.

---

[1] https://developers.google.com/android/reference/com/google/android/gms/location/ActivityRecognitionClient

In this study, we intend to rethink activity awareness through the design and evaluation of a mobile messaging application that introduces activity transition notification and utilizes improved activity recognition technology. Then we provide corresponding design implications.

The contributions of this study are listed as follows:

1) The design and implementation of ActAware, a mobile instant messaging application that integrates activity awareness based on the re-summarized design dimensions and Google Activity Recognition Client.

2) A two-week field study with 31 participants to analyze the user behaviors of ActAware from three aspects, including the usage motivations of activity awareness, users' privacy concerns, and activity uncertainty disclosure.

3) A discussion to present some design recommendations for mobile messaging to better support activity awareness.

The remainder of this paper is organized as follows: Section 2 reviews the related work on context awareness in mobile messaging, especially activity awareness in mobile messaging. Section 3 presents our design and implementation of ActAware. Section 4 presents a two-week field study to evaluate ActAware. Section 5 presents the results of our field study. Section 6 presents the discussion and design recommendations. Section 7 finally summarizes our work.

## 2. Related work

Nowadays, different types of context information are integrated into mobile messaging to increase expressiveness and awareness, including mobile device setting, location, activity, and heart rate. Researchers have studied the user behaviors of context awareness in mobile messaging mainly from the following aspects: the usage motivations of context awareness, users' privacy concerns, context information transitions, and uncertainty in context awareness.

Many studies have developed mobile messaging applications to integrate context awareness and have explored the usage motivations (Tang et al., 2001; Oulasvirta et al., 2005; Liu et al., 2017; Griggio et al., 2019; Jain et al., 2022). Oulasvirta et al. (2005) redesigned the mobile phone's address book and integrated the context awareness of location, ringing profile, whether the phone was being used, etc., in their application. They conducted a field study and found that users took advantage of context awareness to determine the appropriate time to interrupt their friends. Liu et al. (2017) developed a mobile instant messaging application that integrates heart rate awareness and conducted a field study to explore the user behaviors of disclosing heart rate to their friends. They found that users utilized heart rate awareness as a novel and playful form of communication, which increased emotional awareness and chat frequency among one another. Griggio et al. (2019) developed a mobile instant messaging application that visualizes context information between couples, e.g., closeness to home, battery level, steps, and media playing. They conducted a field study and found that context awareness helps couples coordinate and feel more connected with each other. Jain et al. (2022) designed a contextual auto-response messaging agent, which could speculate whether the user is available by analyzing the context information of the user's mobile device settings and automatically sending responses to others. In a two-week field study of 12 users, researchers found that the auto-response agent alleviated users' pressure and obligation to respond, helped users stay focused on important tasks, and reduced users' need to explain unavailability.

Researchers have also studied users' privacy concerns when integrating context awareness in mobile messaging. Some researchers conducted field studies and found that users had privacy concerns about disclosing context information, e.g., location and mobile device settings, with others (Barkhuus et al., 2008; Antila et al., 2011; Jain et al., 2022). Antila et al. (2011) proposed providing more privacy settings for users in mobile messaging. Some researchers used surveys to study users' privacy settings when integrating location awareness in mobile messaging (Consolvo et al., 2005; Patil & Lai, 2005; Khalil & Connelly, 2006). They found that social relationship, the granularity of location, the semantic type of location, etc., affect users' privacy settings for disclosing location in mobile messaging.

Previous studies have found that the transition of context information helps users infer the status of others (Guzman et al., 2007; Ankolekar et al., 2009; Hassib et al., 2017). Ankolekar et al. (2009) highlighted the recently changed mobile device settings in their mobile messaging application Friendlee. Their field study results showed that the highlighted information helps users know the status of others. Hassib et al. (2017) designed HeartChat, a mobile instant messaging application that integrates heart rate awareness. They conducted a field study and found that users were more concerned about the changes in the heart rates of others, as a significant change in heart rate always indicated a transition in physical activity.

Incorrect recognition in context awareness is inevitable, which might affect user experience. Some researchers have studied the impact of hiding uncertainty in context awareness (Anderson et al., 2006; Damián-Reyes et al., 2011). Anderson et al. (2006) developed iSocialize, a mobile messaging application that integrates users' distance and mobile device settings. They did not show the exact distance but the distance range between users. Their field study results showed that using distance range can not only hide distance uncertainty but also protect user privacy. In addition, some researchers have found that disclosing uncertainty is beneficial (Chalmers & Galani, 2004; Antifakos et al., 2004; Lim & Dey, 2011a). Lim et al. (2011a) disclosed the process of awareness and its uncertainty in two context-awareness applications. Their field study showed that disclosing the awareness process and its uncertainty increases user understanding and trust.

In activity awareness, early activity recognition technology could only obtain users' motion information. Bentley et al. (2007) designed a mobile messaging application that allows family members and friends to view each others' motion status, i.e., *moving* or *not moving*, on their mobile phones. Their field study results showed that users could infer others' activity status to plan communication, coordinate in-person get-togethers, and stay connected in each others' lives. In addition, they found that users did not worry about privacy problems and intended to obtain more activity information about others. Recently, improved activity recognition technology has enabled mobile phones to recognize more detailed activity information. In 2017, Google Play Services developed Activity Recognition API, which supports recognizing *still*, *walking*, *running*, *tilting*, *on bicycle*, *in vehicle*, etc. Based on this API, Buschek et al. (2018) developed ContextChat, a mobile instant messaging application that integrates activity awareness. This application only allows users to view each others' activities when sending messages, and the activity information access control is limited. Researchers found that activity is only shown if Activity Recognition API returned an activity with a high confidence value, and only 58.4% of messages displayed activity information. Because of low recognition accuracy, sometimes users could not accurately infer the status of others, and they assumed that the activity is the least interesting context in mobile messaging.

Liu et al. (2021) designed Significant Otter, an Apple Watch application that allows couples to check each others' current status. This application supports context awareness to be turned on and off (when context awareness is turned off, users can manually select their current status; when context awareness is turned on, the application can estimate whether the user is *sleeping*, *walking*, or *running* by using the motion and heart rate information obtained from the Apple Watch). Through a one-month field study of 20 couples, researchers found that users tended to turn on context awareness, which could easily obtain accurate and true information about each other. However, due to the limited information, some users were skeptical of the application's ability to sense activities and hoped that the application could provide more information.

At present, activity recognition technology has been further advanced. In 2020, Google Play Services developed Activity Recognition Client with improved recognition accuracy, less power consumption, and engineering productivity improvement. Activity Recognition Transition API was also developed to detect users' activity transitions. However, the application design and user behaviors of activity transition have not been studied. With the improvement of recognition accuracy, mobile messaging applications that integrate activity awareness can obtain more accurate user activity recognition information. Whether this information raises users' privacy concerns and what are users' motivations for using this information have not been studied. In this study, we intend to design a mobile messaging application that utilizes improved activity recognition technology to explore the usage motivations of activity awareness, users' privacy concerns, and the influence of displaying activity recognition confidence on user trust.

## 3. Design

We re-summarize design dimensions for activity awareness in mobile messaging based on the characteristics of activity information. Then, we conduct a focus group to discuss the potential usage scenarios of activity awareness, privacy problems, and design suggestions for user interfaces. Finally, we describe the design of ActAware, a mobile instant messaging application that integrates activity awareness.

**3.1 Design dimensions**

Researchers have investigated the design dimensions of integrating context awareness into mobile massaging (Oulasvirta, 2008; Ens et al., 2012; Hassib et al., 2017; Buschek et al., 2018). Oulasvirta (2008) designed four mobile messaging applications that integrate context awareness as analytical examples and identified several design dimensions of context information, e.g., modality, level of description, temporal reference, and the amount of context information. Hassib et al. (2017) presented several design dimensions for heart rate awareness, i.e., data representation, sharing trigger, persistence, and granularity. Buschek et al. (2018) proposed a design space for mobile messaging that integrates context awareness, which contains three core dimensions, i.e., context, sharing, and presentation. In activity awareness, the context is the activity obtained from mobile and wearable devices based on activity recognition technology, and the sharing is implicit and continuous. Researchers did not consider the characteristics of activity information in the core dimension of presentation. Therefore, we re-summarize six design dimensions for activity awareness in mobile messaging based on the characteristics of activity information, as shown in Table 1. Each design dimension has 2 or 4 types.

*3.1.1 Activity information abstraction level*

Context information can be described at different levels of abstraction. Hassib et al. (2017) classified the abstraction levels of heart rate into low and high., i.e., low abstraction level means the raw signals (the number of beats per minute), and high abstraction level means the interpreted information (e.g., using colors to reflect the heart rate). Activity information abstraction levels can also be divided into two types, i.e., low and high abstraction levels. Different levels of abstraction reflect the granularity of information. High abstraction level contains coarse-grained information, e.g., *moving* or *not moving*. Low abstraction level contains fine-grained information, e.g., *still*, *walking*, *running*, *on bicycle*, and *in vehicle.*

Table 1. Design dimensions for activity awareness in mobile messaging

| Design dimension | Type | | | |
|---|---|---|---|---|
| Activity information abstraction level | Low | | High | |
| Activity information access control | Person-based | | Conversation-based | |
| Historical activity information disclosure | Current activity | Current activity and its duration | Current activity and historical activities | Current activity, historical activities, and both of their duration |
| Activity information visualization | Text | | Icon | |
| Activity transition notification | Pushing notification | | Not pushing notification | |
| Activity information confidence disclosure | Displaying confidence | | Not displaying confidence | |

*3.1.2 Activity information access control*

Buschek et al. (2018) classified the access to context information in mobile messaging into person-based, conversation-based, and message-based. Person-based means that one's context information can be seen all the time. Conversation-based means that one's context information can only be seen during the conversation. After the conversation ends, context information cannot be accessed by users. If a user wants to get others' context information, she should send a message to start a conversation. Message-based means that one's context information is integrated into each sent message. When users send messages, activity types are limited. Users could send messages when they are *still*,

*walking*, and *in vehicle* (not the driver). However, users are unlikely to be *running* or *on bicycle* when sending messages. Therefore, we do not consider message-based. Activity information access control can be divided into two types, i.e., person-based and conversation-based.

### *3.1.3 Historical activity information disclosure*

Buschek et al. (2018) classified the presentation persistence of context information into ephemeral and persistent. Ephemeral means that the context information is presented in real-time and disappears immediately. For activity information, persistence contains two perspectives, i.e., time (whether the duration of activity is displayed) and information (whether historical activities are displayed). Therefore, historical activity information disclosure can be divided into four types, i.e., current activity; current activity and its duration; current activity and historical activities; current activity, historical activities, and both of their duration.

### *3.1.4 Activity information visualization*

There are various ways to visualize context information, e.g., text, icon, graph, and color (Dey & Guzman, 2006; Antila et al., 2011). Different ways are suitable for representing different context information, e.g., using the line graph to represent heart rate can intuitively reflect its change, and using colors red and green to represent unavailable and available can intuitively reflect one's availability. Activity information contains many different types, and some visual ways, e.g., graphs and colors, are not suitable to intuitively represent this information. Previous researchers usually use text and icons to present activity information (Oulasvirta, 2008; Buschek et al., 2018). Therefore, activity information visualization can be divided into two types, i.e., text and icon.

*3.1.5 Activity transition notification*

Researchers have found that the transition of context information helps users infer the status of others (Guzman et al., 2007; Ankolekar et al., 2009; Hassib et al., 2017). In addition, the Google Activity Recognition Client could detect users' activity transitions. Therefore, activity transition notification can be divided into two types, i.e., pushing notification and not pushing notification.

*3.1.6 Activity information confidence disclosure*

The results of activity recognition are not always correct, which might reduce user trust (Buschek et al., 2018). Previous studies have shown that displaying recognition confidence can reveal the uncertainty of the recognition process, which helps to increase user understanding and trust in the system (Antifakos et al., 2004; Lim & Dey, 2011b). Google Activity Recognition Client provides activity recognition confidence. Therefore, activity information confidence disclosure can be divided into two types, i.e., displaying confidence and not displaying confidence.

**3.2 Focus group**

We conduct a focus group to gather users' discussions and design suggestions of activity awareness in mobile messaging based on re-summarized design dimensions. We are mainly concerned with the following questions:

1) What are the potential usage scenarios of activity awareness in mobile messaging? Do users have privacy problems disclosing their activity information?

2) What are users' preferences for each design dimension when designing a mobile messaging application that integrates activity awareness? What are their design suggestions?

### 3.2.1 Participants

Six participants (referred to as A1-A6) between 25 and 32 years old (average age: 27.5, median: 27, standard deviation: 2.5, female: 3) took part in our focus group through a snowball sampling method. Participants were mainly master and doctoral students of several faculties, including finance, machinery, and computer science. Two participants had experience in designing and developing Apps, and two participants were experts in activity recognition technology. Participants used mobile messaging applications daily, e.g., WeChat, Slack, DingTalk, QQ, and WhatsApp.

### 3.2.2 Procedure

The focus group was conducted online using DingTalk. An author served as the facilitator. First, the facilitator introduced activity awareness to participants to discuss the potential usage scenarios and privacy concerns of activity awareness in mobile messaging. Second, participants were introduced to the re-summarized design dimensions, and they were asked to use these dimensions to design a mobile messaging application that integrates activity awareness. Each participant described their preferences and design suggestions for these dimensions. The facilitator drew prototypes for each design suggestion. Finally, participants discussed these prototypes together and determined the design of six dimensions. The focus group lasted 145 minutes. The entire session was video recorded, transcribed into text, and coded in terms of themes.

### 3.2.3 Focus Group Findings

The focus group results were in Chinese, and the translated versions are presented in this section.

**The potential usage scenarios and privacy concerns of activity awareness**

Participants assumed that activity awareness is extremely beneficial. First, activity awareness can help users infer others' status to decrease interruptions. A5 mentioned, *"If I am driving, I cannot be interrupted, and my friends can use this information to choose the appropriate time to contact me. This function is quite useful."* Second, activity awareness can help to form interest groups and enhance real-time interaction. A2 and A4 reasoned that users may use activity awareness to find people of the same interests and initiate real-time communication. A2 mentioned, *"Groups with the same interests can be established through activity awareness, e.g., cycling and running. With the positioning function, people can know what people around them are doing and promote real-time communication between people with the same interests."* A4 mentioned, *"Distant users with the same sports interests can do sports together through activity awareness."*

Participants expressed their privacy concerns about activity awareness in mobile messaging. A1, A2, and A3 reasoned that persistent activity awareness is not conducive to privacy protection. Mobile messaging applications should allow users to control their activity awareness settings. A1 mentioned, *"I do not mind disclosing activity information with my friends, but I think users who are very concerned about privacy may not willing to disclose activity information persistently."* A2 mentioned, *"It is better to give users control about activity awareness."* A3 mentioned, *"I do not like the feeling that others monitor me all the time, as the privacy is not protected. So I think applications should provide some privacy settings, e.g., the time of awareness, the activity types of awareness, and different activity information for different relationships."*

**Design suggestions for user interfaces**

First, each participant was asked their suggestions for the design of six dimensions.

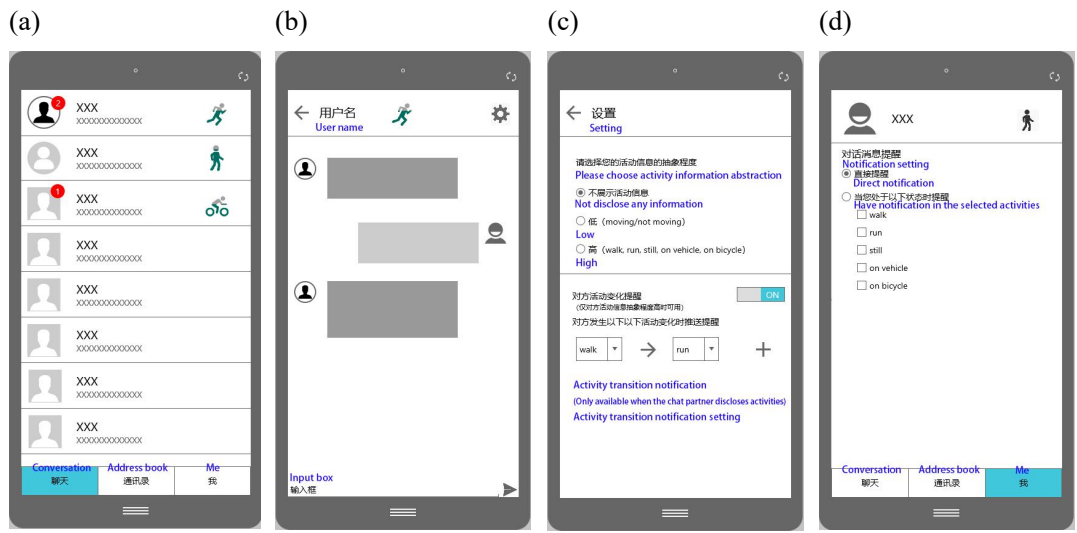

Figure 1. The prototypes discussed in the focus group. (a) Chat list page. (b) Chat page. (c) Detailed page for each chat partner. (d) Setting page.

Based on these suggestions, the facilitator used a quick prototyping software Mockplus[2] to draw prototypes for participants to discuss, as shown in Figure 1. After discussion, the design of six dimensions was finally determined.

For activity information visualization, participants thought icons are better than text, as icons are more intuitive, more interactive, and less interferential. A1 mentioned, *"Icons are more intuitive."* A5 mentioned, *"Too much text on the whole page will disturb users, and icons are relatively less intrusive."*

For activity information abstraction level, participants thought users should have the control to set different abstraction levels of activity information for different relationships. A2 suggested that icons should be designed for the two abstraction levels, i.e., low and high, *"It is better to design separately and let users choose different abstraction levels of activity information for different relationships."*

For activity information access control, participants suggested choosing person-based access control. A3 mentioned, *"Users could check others' activities at any time through person-based access control, which is useful for inferring others' status and*

---

[2] https://www.mockplus.com

*enhancing communication."* Participants also mentioned that person-based access control might cause privacy concerns, and they suggested that the application should provide privacy settings for users. A5 mentioned, *"persistent activity awareness might cause privacy problems, and privacy settings should be provided for users."* For the interface design, participants suggested that activity information should be revealed in both chat lists and chat pages. A4 mentioned, *"Both the chat list and the chat page need to display activities. The activities displayed in the chat list can allow me to quickly judge the status of my friends. The activities displayed on the chat page can help me see my friends' status at any time during the chatting."*

For historical activity information disclosure, participants thought disclosing too much activity information is not conducive to privacy protection, and current activity is sufficient to infer others' status. A1 mentioned, "Current activity is enough." A2 mentioned, *"There is already a lot of information, and any more would be bad for privacy."* A6 mentioned, *"Too much information may cause information overload."*

For activity transition notification, participants thought pushing transition notification is useful, as it can help remind users of others' activity transitions. However, participants were worried that excessive notifications might interrupt users. A2 mentioned, *"This function is useful, but I have received a lot of messages on my phone, and I do not want the application to constantly notify me."* Therefore, A1 and A5 suggested that users should have control over activity transition notifications. A1 mentioned, *"The application needs to allow users to set notification, e.g., the activity types of transition."* A5 mentioned, *"It can be set to notify users once. Users need to set again to continue the notification to avoid frequent messages."*

For activity information confidence disclosure, participants thought displaying confidence is necessary, as the activity recognition result is not always correct. A3 mentioned, *"I think it is necessary to display confidence, and I always wonder whether the recognition results of the system are reliable."* However, participants were concerned that a bad design might cause misunderstanding and information overload in the interface. Therefore, participants proposed different designs for confidence disclosure, as shown in Table 2. Two confidence degrees (30% and 70%) are selected to compare different designs. A1 and A5 reasoned that transparency is not intuitive. A1 mentioned, *"I cannot distinguish the magnitude of confidence."* A5 mentioned, *"There is no comparison."* A6 reasoned that numbers in icons might be too small to see, *"Users cannot discern the numbers."* Double color uses the ratio of the black's height to the icon's height to represent recognition confidence. A2, A3, and A5 reasoned that this design is intuitive. A2 mentioned, *"This design is clear."* A3 mentioned, *"The contrast of this design is obvious."* A5 mentioned, *"Double color design is better, which does not need to add color bar and reduces the interference to users."*

Table 2. Different designs of 30% and 70% recognition confidence disclosure

| Recognition confidence | Transparency | Number | Double color |
|---|---|---|---|
| 30% | 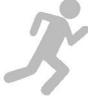 | 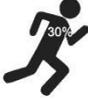 | 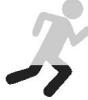 |
| 70% | 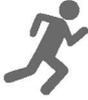 | 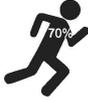 | 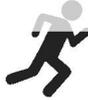 |

### 3.3 ActAware

Based on re-summarized design dimensions and the focus group discussions, we design ActAware, a mobile instant messaging application that integrates activity awareness.

#### *3.3.1 Interface design*

According to the focus group discussions, we only disclose the current activity to users and use an icon to represent the current activity. For activity information abstraction level, we design two sets of icons to represent low and high abstraction levels, respectively, as shown in Figure 2. Except for the *unknown*, the icons of low abstraction level have more details than the icons of high abstraction level. In ActAware, users can choose different abstraction levels of activity information for each chat partner. For activity information access control, we choose person-based access control, and the activity information is displayed on both the chat list page and chat page persistently. For activity information confidence disclosure, in order to explore the influence of displaying confidence on user behaviors, we design two versions, i.e., displaying confidence and not displaying confidence, to conduct a contrast experiment. We use double color to represent activity recognition confidence.

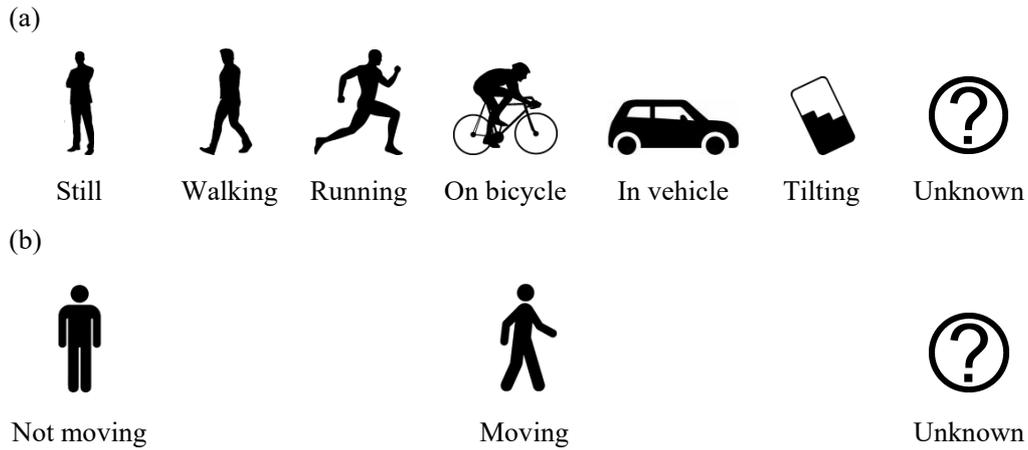

Figure 2. The icons of different activity information abstraction levels. (a) The icons of low abstraction level. (b) The icons of high abstraction level.

We design user interfaces for ActAware. We use Chinese as the primary language in ActAware, which is more convenient for our Chinese participants. The translated version is shown in Figure 3. On the chat list page, users can find their current activities on the top of the chat list and chat partners' current activities on the right of each one's chat row (Figure 3a). The icons in the prototypes (Figure 1a) were larger than the icons in the actual application (Figure 3a). In the actual application, the chat list has to display the specific time of chatting, so the space left for the icon is small. On the chat page, users can see a chat partner's activity at the top of the page (Figure 3b). There is a setting icon at the top-right corner of the chat page. After clicking the setting icon, users can configure the activity information abstraction level and activity transition notification for each chat partner. Users cannot control their chat partners' activity transition notification, nor can they know what activity transition notification their chat partners have set for them (Figure 3c). Users can set the hidden time period (others cannot see any type of activities during a certain time period) and the hidden activity type (others cannot see certain types of activity) on the setting page (Figure 3d).

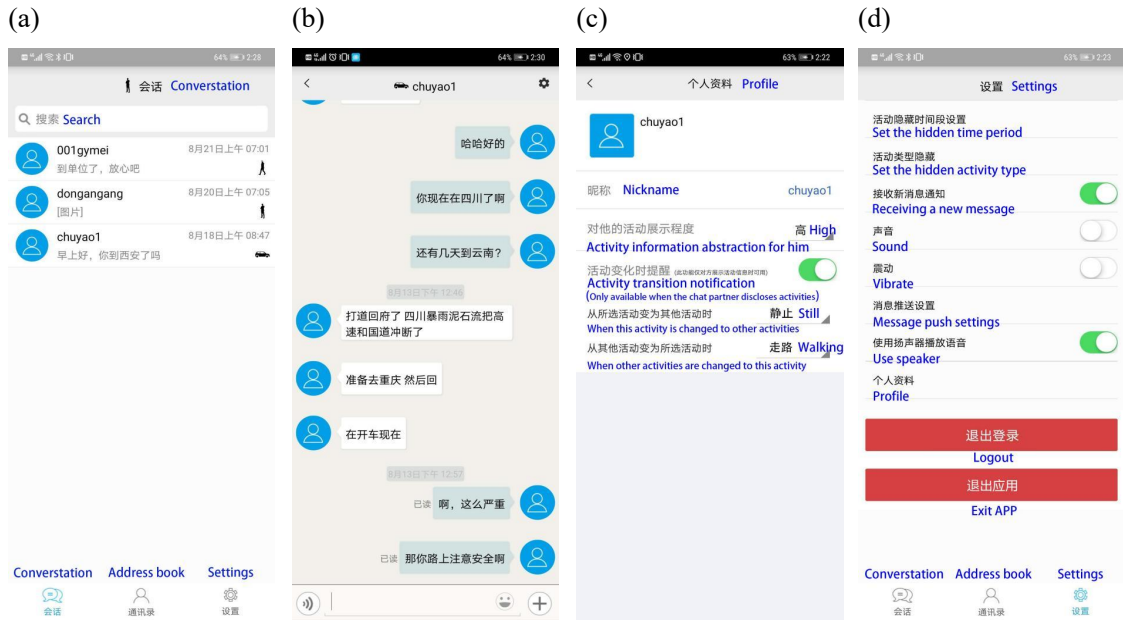

Figure 3. The user interfaces of ActAware. (a) Chat list page. (b) Chat page. (c) Detailed page for each chat partner. (d) Setting page.

### 3.3.2 Implementation

ActAware is implemented as an Android application, which utilizes the Easemob instant messaging SDK[3] and a MySQL database. Activity information is recognized by Google Activity Recognition Client, which offers the following activities: *in vehicle*, *on bicycle*, *running*, *walking*, *still*, and *tilting* (the mobile device is being lifted and is having some angle with the flat surface, which usually means that users pick up phones and use phones). ActAware is set to read sensor data every 10 seconds and generates a record if activity changes. The database stores users' activity records, activity transition notifications, and activity setting records.

---

[3] https://www.easemob.com/product/im

## 4. Evaluation

We conduct a two-week field study to evaluate ActAware. We ensure that participants know the purpose of the field study and the confidentiality of the data. All of the collected data were only used for this study and were not disclosed in public.

### 4.1 Participants

Thirty-one Chinese participants (12 groups, referred to as P1-P31) volunteered to take part in the study through a snowball sampling method. Participants in each group used Android phones and communicated with each other at least once daily. Participants ranged in age from 21 to 63 years (average: 30.6, median: 27, standard deviation: 11.4, female: 9). Participants included seventeen students, eleven full-time job workers, and three retirees with different backgrounds, e.g., education, computer science, and architecture. As shown in Table 3, the social relationship included friends, family, and couples.

### 4.2 Procedure

After signing the consent form, participants installed ActAware and added chat partners. Participants could only have a single chat instead of a group chat with chat partners in ActAware. Moreover, participants could not use ActAware to chat with people other than chat partners, as only recruited participants were observed during the study. ActAware had two versions, i.e., displaying confidence and not displaying confidence. Participants were randomly divided into two crowds of similar sizes. Group 1-5 (15 participants) installed the displaying confidence version, and Group 6-12 (16 participants) installed the not displaying confidence version. A face-to-face meeting or a video call was conducted with each participant. Participants were introduced to the concept of activity awareness, the functions of ActAware, and field study requirements, i.e., participants in a group were required to contact each other only through ActAware during the study. Participants had been told that their chat partners could use activity transition notification to track their activities. Then participants were asked questions on demographic information.

Table 3. Participants

| Group | Participant | Age | Sex | Social Relationship |
|---|---|---|---|---|
| 1 | P1 | 28 | Female | Family |
|   | P2 | 55 | Female |  |
|   | P3 | 58 | Male |  |
|   | P4 | 30 | Male |  |
|   | P5 | 57 | Female |  |
| 2 | P6 | 29 | Female | Friend |
|   | P7 | 29 | Male |  |
|   | P8 | 28 | Male |  |
| 3 | P9 | 23 | Male | Couple |
|   | P10 | 23 | Female |  |
| 4 | P11 | 24 | Male | Couple |
|   | P12 | 24 | Female |  |
| 5 | P13 | 23 | Male | Friend |
|   | P14 | 23 | Male |  |
|   | P15 | 23 | Male |  |
|   | P16 | 24 | Male |  |
| 6 | P17 | 31 | Male | Friend |
|   | P18 | 25 | Male |  |
|   | P19 | 25 | Male |  |
| 7 | P20 | 21 | Male | Couple |
|   | P21 | 23 | Female |  |
| 8 | P22 | 29 | Female | Couple |
|   | P23 | 36 | Male |  |
| 9 | P24 | 27 | Male | Friend |
|   | P25 | 30 | Male |  |
| 10 | P26 | 29 | Female | Friend |
|   | P27 | 25 | Male |  |
| 11 | P28 | 24 | Male | Friend |
|   | P29 | 24 | Male |  |
| 12 | P30 | 37 | Male | Family |
|   | P31 | 63 | Male |  |

During the two-week field study, participants were asked only to use ActAware to communicate with each other. After the first week of usage, participants were asked to rate their trust in activity recognition results. Then Group 1-5 and Group 6-12 exchanged their versions.

After the field study, participants were asked to complete a questionnaire and an interview to collect their feedback on ActAware. Previous work on context awareness in mobile messaging mainly focused on three aspects, i.e., the usage motivations of context awareness, users' privacy concerns, and uncertainty in context awareness (Bentley & Metcalf, 2007; Lim & Dey, 2011a; Hassib et al., 2017; Buschek et al., 2018). Therefore, our questionnaire and interview also focused on the above three aspects. The questionnaire mainly investigated whether ActAware can help users judge the status of others, whether the activity transition notification is useful, whether users have privacy concerns, whether the privacy settings are useful, and whether users trust the activity recognition results. The interview mainly investigated the usage scenarios of activity transition notification, whether users have privacy concerns and reasons, the motivations for using ActAware, and users' attitudes about displaying confidence. Each interview lasted about 30 minutes. Details of the questionnaire and interview are shown in Appendix.

## 5. Results

We present the results from data logs, questionnaires, and interviews. The results of questionnaires and interviews are presented from three aspects, including the usage motivations of activity awareness, users' privacy concerns, and recognition confidence disclosure. Considering the small sample size, we do not analyze the differences between the two-person and multi-person groups, nor do we analyze the influence of different social relationships on user behaviors.

### 5.1 Data logs

During the two-week field study, 137,073 activity records were collected. P30 only had 8 activity records. Therefore, we exclude the results of Group 12 (P30 and his chat partner P31). The rest of twenty-nine participants had 131,275 activity records (average: 4,526, min: 558, max: 16,553). Table 4 shows the activity types and frequencies. Each activity record collects the recognition confidence, and the average confidence is 83.5% (median: 94%, mode: 100%, standard deviation: 21.2). Totally 36,051 activity records have 100% recognition confidence.

We collect users' activity setting records, which include the settings of activity information abstraction level, hidden time period, and hidden activity type. We found that most participants used default settings during the study. Only P15 and P17 changed their activity information abstraction level from low to high for chat partners.

We collect 162 activity transition notifications, including changes from a certain activity to other activities, and changes from other activities to a certain activity, as shown in Table 5. In this table, we also show the number of participants setting the notification for each activity type. Totally twenty participants (68.9%) used this notification.

### 5.2 Questionnaires

After the field study, participants were asked to use a 5-point Likert scale to complete a questionnaire. We use a one-sample t-test to testify the significance between the collected scores and the neutral point (3). If $p<0.05$, it means that the results are statistically significant. The results of the questionnaires are shown below.

Table 4. Activity types and frequencies

| Activity type | Frequency | Percentage |
| --- | --- | --- |
| Tilting | 58,517 | 44.6% |
| Still | 51,047 | 38.9% |
| In vehicle | 10,810 | 8.2% |
| Walking | 8,927 | 6.8% |
| On bicycle | 1,047 | 0.8% |
| Running | 299 | 0.2% |
| Unknown | 628 | 0.5% |

Table 5. Activity transition notifications

| Activity type | From this activity to other activities (Frequency and percentage) | The number of participants setting this type of notification | From other activities to this activity (Frequency and percentage) | The number of participants setting this type of notification |
| --- | --- | --- | --- | --- |
| In vehicle | 51 (31.5%) | 13 | 48 (29.6%) | 12 |
| Still | 48 (29.6%) | 11 | 57 (35.2%) | 12 |
| Walking | 25 (15.4%) | 8 | 28 (17.3%) | 7 |
| Running | 16 (9.9%) | 5 | 19 (11.7%) | 6 |
| On bicycle | 11 (6.8%) | 3 | 5 (3.1%) | 2 |
| Tilting | 4 (2.5%) | 2 | 2 (1.2%) | 1 |
| No notification | 7 (4.3%) | 2 | 3 (1.9%) | 2 |

Participants reasoned that ActAware is useful for inferring others' status (score: 3.86, $t=5.88$, $p<0.001$).

Participants had low privacy concerns about activity awareness (score: 2.41, $t=-2.82$, $p<0.01$), and they reasoned that activity settings could reduce privacy concerns (score: 3.83, $t=5.26$, $p<0.001$).

Participants reasoned that activity transition notification is useful (score: 3.83,

$t$=5.26, $p$<0.001).

Participants trust the activity recognition results in both the displaying confidence version and the not displaying confidence version (the displaying confidence version score: 3.90, $t$=5.36, $p$<0.001; the not displaying confidence version score: 3.65, $t$=4.89, $p$<0.001).

We use a contract experiment to explore the influence of displaying confidence on user trust. First, we use an independent-sample t-test and find that there is no significant difference between the scores of Group 1-5 and Group 6-11, which excludes the effect of version order. Then we use a paired-sample t-test and find that there is no significant difference between the scores of the displaying confidence version and the not displaying confidence version. These results indicate that whether to display confidence does not affect user trust in our study.

### 5.3 Interviews

All of the interviews were transcribed into words and were coded in terms of three themes, including the usage motivations of activity awareness, users' privacy concerns, and recognition confidence disclosure. In these three themes, grounded theory was adopted to establish new ideas emerging in each theme as sub-themes, until no more sub-themes could be found in the interviews (Aronson, 1995; Charmaz & Belgrave, 2012). The results are listed as follows.

#### *5.3.1 The usage motivations of activity awareness*

Participants could use the persistently displayed activity information or activity transition notification to check others' activities. Participants said that they use the persistently displayed activity information to speculate on the reasons for chat interruption, plan communication, speculate on whether chat partners are departing/arriving, and deepen the understanding of chat partners' living patterns. Participants also mentioned that activity transition notification could help them plan communication, speculate on whether chat partners are departing/arriving, and deepen the understanding of chat partners' living patterns.

### Speculate on the reasons for chat interruption

Thirteen participants (44.8%) said they use the persistently displayed activity information to speculate on the reasons for chat interruption. Participants mentioned that they do not expect chat partners to reply in time when chatting. They usually send messages to chat partners and wait until they get replied. When the chat partner fails to reply in time, they speculate the reasons for the delay through the persistently displayed activity information and wait for the chat partner's reply. P27 mentioned, *"One day I checked her activities (P26) and found that she was in vehicle. I could infer that she was driving, so I knew the reason why she did not reply."* Moreover, they could use the persistently displayed activity information to determine the reasons for the non-reply during chatting. P1 mentioned, *"One day P4 and I were chatting, and he suddenly stopped replying. I noticed that his activity changed to walking, and I speculated that he might be suddenly busy."*

### Plan communication

Ten participants (34.5%) said they usually check chat partners' activities first when they intend to plan communication. If the activity is *walking* or *in vehicle*, it usually means that the chat partner is busy. The participant would choose another time to contact the chat partner. P19 said, *"One day I wanted to ask P17 some questions, and I found he was walking, so I guessed he might not reply to me now, and I decided to ask him later."* P18 mentioned, *"I find my chat partner usually watches his phone when his activity is tilting, so I choose this time to contact him."*

Five participants (25.0%) indicated that they could easily obtain chat partners' activity transitions through activity transition notification and find the appropriate time to contact them. P1 set the notification several times when P2's activity changed from other activities to *walking* or *in vehicle*, *"Sometimes I want to chat with P2, and I know she is available when taking a bus or walking home. So I set a notification to find the right time to call her."* P14 set the notification when P13's activity changed from *on bicycle* to other activities and from other activities to *tilting*, *"One day I wanted to chat with P13. I saw he was riding a bike, and I set a notification to remind me. I also find that tilting usually means my chat partner is using his cell phone, so I set the notification to chat with him at this time. He may reply to me more quickly."*

**Speculate on whether the chat partner is departing/arriving**

Six participants (20.7%) said they use the persistently displayed activity information to speculate on whether chat partners are departing/arriving before meeting each other. P9 said that she eats with P10 at noon every day, and they meet at the restaurant. She speculated on whether P10 had set off by using activity awareness, *"If I see his activity is walking, it means that he has already set off."* P4 mentioned judging whether P1 had arrived at the meeting place, *"One day we were going to meet at a place, and I wanted to check if she was there. I saw her activity changed from in vehicle to walking, then I figured that she was almost there."*

Eight participants (40.0%) indicated that activity transition notification could help to speculate on whether chat partners are departing/arriving before meeting each other. P9 set the notification when P10's activity changed from other activities to *walking*, *"One day I wanted to eat with him. When his activity was changing to walking, I knew he had already set off."* P22 set the notification when P23's activity changed from other activities to *in vehicle* and from *in vehicle* to other activities. She mentioned that she could receive the notification in the mobile phone notification bar, which helped her to directly check her husband's activity transitions, *"My husband drives back home every night, so I set the notification to remind me that he is driving home. I do not need to open ActAware, and I can know his activity transitions in the mobile phone notification bar. I usually begin cooking dinner when I receive the notification. I also set the notification when his activity changes from in vehicle to other activities, and the received notification can let me know that he is almost arriving."*

**Deepen the understanding of living pattern**

Eleven participants (37.9%) reported using the persistently displayed activity information to learn chat partners' living patterns and routines to increase closeness. P2 stated that she usually looks at her daughter P1's activities to learn P1's living routines, *"My daughter is living far from me, and I am concerned about her. I want to know her living routines to determine if she is working overtime or going home late at night."* She felt *"more closely"* with her daughter. P11 and P12 are in a long-distance relationship. P11 said, *"Although we usually chat and know each others' living patterns, I still want to look at P12's activities to learn more."*

Five participants (25.0%) indicated that they could deepen their understanding of chat partners' status through activity transition notification. P18 set a notification when P19's activity changed to *tilting*, which indicated that P19 was available. P18 mentioned that he set the notification not to plan communication, but to know P19's routines, *"Tilting generally means that he picks up and uses his cell phone, which means that he is not busy according to my prior knowledge about him. I can know his schedule through this notification."* P11 and P12 are in a long-distance relationship, and they said that by using activity transition notification, they could know more about each other. P11 said, *"We usually contact each other at a fixed time in the morning and evening. I*

*do not fully know what he is doing during the rest of the time, and I am too busy to check his status from time to time. Activity transition notification can let me know more about him."*

*5.3.2 Users' privacy concerns*

Participants were asked whether they have privacy concerns and their reasons. The results showed that 27 participants do not have privacy concerns. The reasons for not having privacy concerns were extracted and analyzed.

**The activities in ActAware contain limited information**

Twenty-one participants (72.4%) mentioned that the activities displayed in ActAware are mainly simple activities, which contain limited information. P18 reasoned that location is more sensitive than the activities in ActAware, *"This information will not reveal my exact location, so I do not have any privacy concern."* P19 reasoned that this information is not enough for acquaintances to infer his specific state, *"I usually worry about acquaintances knowing my specific state, but this activity information is not enough for them to figure out exactly what I am doing, so I do not have any privacy concerns."* P20 mentioned that he does not have privacy concerns. Instead, he wanted more activity information, *"P21 works at home, and her activities are usually still or tilting. Tilting may indicate that she is using her mobile phone, but I cannot judge whether she is using her mobile phone to work or play games. So I need more information."*

**Close relationship**

Eighteen participants (62.1%) mentioned that chat partners are in a close relationship with them. Although chat partners could use activity information to infer participants' specific states, e.g., the chat partner could use prior knowledge to infer the participant is driving instead of taking a bus through *in vehicle*, participants trust chat partners. They do not worry about chat partners abusing their information. P2 said, *"I do not have privacy concerns because they are my family, and I trust them very much."*

**Activity privacy settings**

Eleven participants (37.9%) said that activity privacy settings in the application alleviate their privacy concerns to some extent. Participants said that they would adjust activity privacy settings according to the change of chat partner, time, and usage scenario. Activity privacy settings ensure participants adjust at any time, which helps protect their privacy. P20 said, *"Activity privacy settings are necessary because I can control my information. Although I do not use these settings during the experiment, I may use them in the future."*

Only two participants (6.8%) expressed their privacy concerns. P15 and P17 adjusted the activity abstraction level from low to high during the study. Both P15 and P17 indicated that they are very privacy-conscious, *"I do not want my friends to know more"*, and *"I want to hide more than the existing settings, which I think will better protect my privacy."* In addition, P15 and P17 reasoned that activity privacy settings are helpful, and reduce their privacy concerns to some extent.

*5.3.3 Recognition confidence disclosure*

Participants were asked about the influence of displaying confidence on trust. Twenty-one participants (72.4%) reported that by checking their activities and asking for chat partners' activities, they found that the recognition accuracy is high, so they trust these results. During the study, the recognition confidence was often at a higher level, so participants reasoned that whether to display recognition confidence did not affect their judgment. P2 said, *"Every morning P3 drives to work. I often check his activity and find that the recognition result is always correct. So I trust the activity recognition results. I do not think it is necessary to display recognition confidence."* P20 said, *"I trust the*

*activity recognition results, because I find that my chat partner's activity recognition results are always correct."*

Eight participants (27.6%) mentioned that they encountered inaccurate activity recognition results during the study. They reasoned that inaccurate activity recognition results would reduce user trust. At this time, displaying confidence could help users understand the system, reduce the negative impact of inaccurate recognition, and improve user trust. P22 said, *"One day I was walking on the road, and I saw that my activity was in vehicle with low confidence. I walked for a while, and I saw my activity change to walking. Maybe there was a certain delay in the recognition process, and displaying confidence made me feel better."* P6 said, *"Displaying confidence is useful when the activity recognition result is inaccurate, which could let me think that the system is not completely wrong."* Four participants also reasoned that only displaying confidence is not enough when the confidence is low. The system needs to show more detailed activity recognition results and even the principle of activity recognition. P21 said, *"The system needs to show more information, and only displaying confidence is not enough."* P8 said, *"I do not understand the activity recognition technology, and I need to know more about the activity recognition process. It is best to tell me the principle of activity recognition."*

## 6. Discussion and design recommendations

We discuss the research results from the aspects of activity transition notification, usage motivation, and users' privacy concerns. Then we provide design recommendations for mobile messaging applications to better support activity awareness.

### 6.1 Activity transition notification

Researchers have found that the transition of context information is helpful for users to infer the status of others (Guzman et al., 2007; Ankolekar et al., 2009; Hassib et al., 2017). Ankolekar et al. (2009) found that the recently changed mobile device settings are helpful for users to determine the status of others. Hassib et al. (2017) found that users were more concerned about the changes in others' heart rates. We expand the results of previous work, introduce activity transition notification in activity awareness for the first time, and investigate its impact. We found that activity transition notification is useful for planning communication, speculating on whether the chat partner is departing/arriving, and deepening the understanding of living patterns. Furthermore, activity transition notification supports users to view chat partners' activity transitions in the mobile phone notification bar instead of open ActAware, which conveniences users to some extent.

According to Table 4 and Table 5, the activity transition notifications of *still* and *in vehicle* take the largest part, which may be because these two activities constantly occur in daily life and help users achieve various motivations. *In vehicle* means that a user drives or rides in a car. Setting the notification from other activities to *in vehicle* or from *in vehicle* to other activities may help users speculate on whether chat partners are departing/arriving before meeting each other. Furthermore, when a user is driving, it is not suitable for interruption, and setting the notification from *in vehicle* to other activities may help users plan communication. Most of the activities that the system recognized are locomotion activities, e.g., *walking*, *running*, *in vehicle*, and *on bicycle,* which may not suitable for interruption. Activity transitions from other activities to *still* may mean that the user has arrived at an office or arrived at home after work. Setting such notifications may help users plan communication and deepen their understanding of chat partners' living patterns. *Tilting* usually means that a user picks up the mobile phone and uses it, which occurs most frequently. However, it is challenging to determine whether the user is available for interruption based on the information of picking up the mobile phone and using it (the use of mobile phone maybe for work or not). Therefore, users are less likely to set the activity transition notification of *tilting*. Although the activity transitions of *running* and *on bicycle* can help users infer chat partners' specific status, they less occur in daily life and users pay less attention to them.

**6.2 Usage motivation**

Researchers have studied the usage motivations of context awareness in mobile messaging applications (Oulasvirta et al., 2005; Bentley et al., 2007; Liu et al., 2017; Griggio et al., 2019). Context includes location, motion, heart rate, and mobile device settings. Oulasvirta et al. (2005) conducted a field study in a mobile phone's address book application that integrated the context awareness, e.g., location, ringing profile, and whether the phone was being used. They found that users took advantage of context awareness to determine the appropriate time to interrupt their friends. Bentley et al. (2007) designed a mobile messaging application that allows family members and friends to view each others' motion status. Their field study results showed that users could use motion awareness to plan communication, try to arrive at the same place at the same time, and increase the feelings of connectedness. Liu et al. (2017) developed a mobile instant messaging application that integrates heart rate awareness and conducted a field study to explore the user behaviors. They found that users utilized heart rate awareness as a novel and playful form of communication, which increased emotional awareness and chat frequency among one another. These researchers have found that the usage motivations of context awareness included planning communication, coordinating in-person get-togethers, and staying connected in each others' lives. In our study, the usage motivations of activity awareness include speculating on whether the chat partner will reply in time, planning communication, deepening the understanding of living patterns, and speculating on whether the chat partner is departing/arriving. Our results verify and expand the conclusions of previous work. We found that both activity awareness and other context awareness in mobile messaging are essential to help users understand each others' current states, meet their needs to contact each other, and enhance emotional interaction.

Furthermore, the purposes of these usage motivations can be divided into two categories, i.e., practical and emotional purposes. The practical purpose usually means that users view each others' activities for some practical reasons, e.g., speculating on whether the chat partner will reply in time, planning communication, and speculating on whether the chat partner is departing/arriving. The emotional purpose usually means that users view each others' activities out of emotional needs. Users want to know each others' living patterns and get closer.

In these two categories, the practical purpose tends to emphasize more on timeliness. Users utilize activity awareness to find the appropriate time to initiate real-time communication (including face-to-face communication and online communication) with others. The emotional purpose may not have a high requirement on timeliness. Users may not initiate real-time communication, but to better understand each other and maintain relationships, which is consistent with the concept of phatic interactions (Vetere et al., 2009; Buschek et al., 2018), namely the act of context exchange itself is beneficial for users to maintain human relationships.

In addition, we speculate that context information access control might have an effect on usage motivations. Context information access control could be classified into person-based, conversation-based, and message-based (Buschek et al., 2018). Previous work studied the usage motivations of context information in conversation-based and message-based access control (Cho et al., 2020; Jain et al., 2022). Cho et al. (2020)

designed MyBulter, which discloses the user's context information only when the sender has sent a message. Jain et al. (2022) designed a contextual auto-response messaging agent, which could send auto-response messages by modeling availability using smartphone sensors. Researchers found that context information could help alleviate users' pressure to respond and reduce users' need to explain unavailability in conversation-based and message-based access control. These usage motivations emphasize more on the practical purpose. In our study, we explore the usage motivations of activity awareness in person-based access control. Person-based access control supports users to view others' activity information persistently. Except for the practical purpose, we found that persistent activity awareness could help deepen the understanding of others' living patterns.

## 6.3 Users' privacy concerns

Researchers have studied users' privacy concerns about mobile messaging applications that integrate context awareness and have found that users had privacy concerns about disclosing their location, mobile device settings, and other contexts (Barkhuus et al., 2008; Antila et al., 2011). However, users do not worry about privacy problems of disclosing their motion information, and hope to obtain more activity information. These results indicated that the amount of information contained in context might have a specific impact on users' privacy concerns. Users have privacy concerns in mobile messaging applications that integrate a variety of contexts, as these contexts contain too much information. On the contrary, compared with disclosing other types of context information, users have fewer privacy concerns about disclosing motion in the mobile messaging application, as motion is a simple and limited context.

  We expand the results of previous work and investigate users' privacy concerns when integrating simple activity in mobile messaging. Although activity awareness in

ActAware supports obtaining more types of activities and contains slightly more information than motion awareness, these simple activities are still a simple and limited context, and users' privacy concerns are not significantly increased.

We found that the method of context information collection might have an effect on user privacy. Previous work designed mobile messaging applications that support users to manually select their status (Cho et al., 2020; Liu et al., 2021). Compared with automatic awareness, the manual setting could give users absolute control over the disclosed information and help protect user privacy. However, manual setting leads to new problems, i.e., the burden of manual setting and plausible deniability. Therefore, mobile messaging applications need to find a balance between automatic context information collection and user privacy.

Relationships have a significant effect on disclosing context awareness in mobile messaging, and researchers found that users were more willing to disclose context information with close relationships (Knittel et al., 2013; Jain et al., 2021). Our study results lead to the same conclusions of previous work. Although we focus on close relationships, we could speculate that participants' attitudes about activity awareness are conservative. Participants in the field study mentioned that they do not worry about chat partners abusing their information because they are in a close relationship, and they may adjust activity privacy settings on change of chat partner, time, and usage scenario. Therefore, we could speculate that participants might be cautious about disclosing activity information to not-so-close relations.

## 6.4 Design recommendations

### 6.4.1 Integrating more types of information

Existing mobile messaging applications that integrate activity awareness could recognize a few types of simple activities, which contain limited information to help users. We found that the activity information of *tilting* is not enough for users to infer whether chat partners could be interrupted. Simultaneously, we found that disclosing simple activity does not cause significant privacy concerns, so we suggest mobile messaging applications integrate more types of simple activities when technology permits to help users better infer each others' status. In addition, inspired by previous work (Cho et al., 2020; Liu et al., 2021), we suggest mobile messaging applications support the manual setting. Users could choose to manually set or use applications to automatically be aware of their activity information to chat partners.

### *6.4.2 Protecting user privacy*

Although disclosing simple activity does not cause significant privacy concerns, it is still necessary to protect user privacy, and privacy settings are essential. In addition, considering that the usage scenarios will continue to change, we suggest mobile messaging applications introduce a privacy nudge (Zhang & Xu, 2016; Gabriele & Chiasson, 2020), which prompts users to reflect on whether they need to adjust their privacy settings at appropriate times. In activity transition notification, mobile messaging applications should also protect user privacy. We suggest mobile messaging applications introduce corresponding control, e.g., activity transition notifications set by users can only take effect with the consent of chat partners.

### *6.4.3 Improving user understanding*

Previous studies have shown that improving user understanding of the recognition process can increase user trust in the recognition results (Lim & Dey, 2011b; Dong et al., 2017). Our results show that displaying confidence when the activity recognition result is inaccurate can help users understand the system and reduce the negative influence.

Therefore, we suggest mobile messaging applications display confidence according to its value. When the confidence is high, it is not necessary to display. However, when the confidence is low, the activity recognition result is probably inaccurate, and displaying confidence can improve user understanding and trust. Furthermore, our results show that displaying confidence is not enough when the confidence is low. Therefore, we suggest mobile messaging applications introduce detailed information about the activity recognition process, e.g., the recognition confidence of each activity (not just the recognized activity).

## 7. Conclusions and future work

In this study, we rethink activity awareness through the design and evaluation of a mobile messaging application that utilizes improved activity recognition technology. First, we re-summarize six design dimensions of activity awareness based on the characteristics of activity information. Second, we design and implement ActAware, a mobile instant messaging application that integrates activity awareness based on the re-summarized design dimensions and Google Activity Recognition Client. Third, we conduct a two-week field study to evaluate ActAware and fully explore user behaviors from three aspects, including the usage motivations of activity awareness, users' privacy concerns, and recognition confidence disclosure. The results show that activity awareness can help users speculate on the reasons for chat interruption, plan communication, speculate on whether the chat partner is departing/arriving, and deepen the understanding of living patterns. Compared with disclosing other types of context information, users have fewer privacy concerns about disclosing their simple activities in ActAware. The results also show that whether to display confidence does not affect user trust, as the recognition confidence was often at a higher level during the study. Finally, we provide design recommendations for mobile messaging applications, i.e., integrating more types of information, protecting user privacy, and improving user understanding.

The limitations of our work are that we have only six graduate participants in the focus group and a small sample size of twenty-nine participants (including only one group of family and four pairs of couples) in the field study. There might exist some bias in our findings, as we do not consider the requirements of people other than graduate students in the focus group and the usage scenarios of activity awareness between other social relationships in the field study. Our designed App was lack of a follow-up validation before the field study, so we cannot be sure that the ideas developed through the focus group are accurately represented in the design. Furthermore, for the design dimensions of activity information access control, historical activity information display, and activity information visualization, we only evaluate the person-based, current activity, and icons. We do not consider other types. In addition, our participants are all Chinese users, and we do not address the effects of cultural differences.

In the future, we intend to explore other types of design dimensions and corresponding user behaviors, e.g., the usage scenarios and users' privacy concerns about historical activities. We also intend to conduct a fine-grained study by focusing on specific groups and relationships, e.g., long-distance couples, parents and children, and fitness people, to explore user behaviors and usage scenarios.

*Computer Supported Cooperative Work*, 1676-1690. https://doi.org/10.1145/2818048.2820073

# Appendix

**Questionnaires after the field study**

Please rate the following questions:

1. Do you think ActAware is useful for inferring others' status?

Very useless  ○1  ○2  ○3  ○4  ○5  Very useful

2. Do you have privacy concerns about disclosing your activity information to your chat partner?

Definitely not concern ○1  ○2  ○3  ○4  ○5  Definitely concern

3. ActAware allows users to set activity information abstraction level, hidden time period, and hidden activity type. Do you think these settings could reduce your privacy concerns?

Definitely not reduce  ○1  ○2  ○3  ○4  ○5  Definitely reduce

4. Do you trust the activity recognition results of your chat partner?

Fully distrust  ○1  ○2  ○3  ○4  ○5  Fully trust

5. Do you think activity transition notification is useful?

Very useless  ○1  ○2  ○3  ○4  ○5  Very useful

**Interview questions after the field study**

1. How do you infer the other's current status (e.g., commuting and working) based on the activity information provided by ActAware?

2. Under what scenarios do you look at other's activities? What are the motivations for looking at it?

3. Did you change the settings of activity abstraction level/hidden time/hidden activity type during the field study? If so, how did you set them? Why did you set them? If not, why did you not change these settings?

4. Under what scenarios did you use activity transition notification? What is the motivation for using it?

5. What do you think could be improved about activity transition notification?

6. What is the use of confidence information for you? How do you use this information specifically?

Thanks for your answer!